\documentstyle[aps,preprint]{revtex}
\begin{document}
\draft
\gdef\journal#1, #2, #3, 1#4#5#6{{#1~}{#2} (1#4#5#6) #3}
\gdef\ibid#1, #2, 1#3#4#5{{#1} (1#3#4#5) #2}
\def\xhi{{\raise.35ex\hbox{$\chi$}}}
\def\ve{\varepsilon}
\def\eps{\epsilon}
\def\nd{^{\vphantom{\dagger}}}
\def\yd{^\dagger}
\def\frac#1#2{{\textstyle{#1 \over #2}}}
\def\half{\frac{1}{2}}
\def\third{\frac{1}{3}}
\def\fourth{\frac{1}{4}}
\def\intl{\int\limits}
\def\cB{{\cal B}}
\def\eg{{\it e.g.\/}}
\def\etal{{\it et al.\/}}
\def\kT{k_{\scriptscriptstyle{\rm B}}T}
\def\kB{k_{\scriptscriptstyle{\rm B}}}
\def\bt{{\tilde b}}

\preprint{SHS-96-1}
%\twocolumn[\hsize\textwidth\columnwidth\hsize\csname @twocolumnfalse\endcsname

\title{Thermodynamics for Fractional Exclusion Statistics}
\author{ Serguei B. Isakov$^{a}$, Daniel P. Arovas$^b$,  Jan Myrheim$^{c}$,
Alexios P. Polychronakos$^{d}$\\  }
\address{Centre for Advanced Study, Norwegian Academy of Science and Letters,\\
Drammensveien 78, 0271 Oslo, Norway}
\address{and}
\address{${}^{a}$Medical Radiological Research Center, Obninsk,
Kaluga Region 249020, Russia}
\address{${}^b$Physics Department 0319, University of California at San Diego,
La Jolla CA 92093, USA}
\address{${}^{c}$Department of Physics, University of Trondheim,
N--7034 Trondheim, Norway}
\address{${}^{d}$Theoretical Physics Department, Uppsala University,
Box 803, S--75108 Uppsala, Sweden \\ }

\date{January 23, 1996}

\maketitle

\begin{abstract}
We discuss the thermodynamics of a gas of free particles obeying Haldane's
exclusion statistics, deriving low temperature and low density expansions.  
For gases with a constant
density of states, we derive an exact equation of state and find
that temperature-dependent quantities are independent of the statistics parameter.
\end{abstract}

\bigskip

%\vskip2pc]

\narrowtext

Haldane's definition of fractional statistics \cite{Hal1} is based on the
generalized exclusion principle $\Delta d(N) = -g\Delta N$, where $d(N)$ is
the dimension of the single particle Hilbert space when $N-1$ states are
occupied and $g$ is the exclusion statistics parameter ($g\nd_{\rm Bose}=0$,
$g\nd_{\rm Fermi}=1$).  
This leads to a statistical weight
\begin{equation}
W={[K-(g-1)(N-1)]!\over N!\,[K-g(N-1)-1]!} \,; \quad K=d(1)
\label{weight}
\end{equation}
which, when applied locally in phase space ({\it i.e.}\  to a set of
neighboring energy levels), results in distribution functions for
free particles obeying fractional exclusion statistics (FES) \cite{I1,Wu,R}.
Corresponding thermodynamic properties have been discussed \cite{Wu,NW}.
Remarkably, the single-state grand partition functions for FES previously 
appeared  as describing the thermodynamics of anyons in the 
lowest Landau level \cite{dVO1}, and  
connection with FES was then established \cite{Wu}.  
The FES distributions were also derived
\cite{I2,BW,Ha,MS} for the elementary excitations of the Calogero \cite{Cal}
and Sutherland \cite{Suth} models; the latter models may be viewed
as systems with purely ``statistical'' interactions \cite{LM1,Poly1}.  
FES is sometimes called {\it ideal} FES to distinguish it from  
exclusion statistics with nonlocal statistical interactions related to 
more general integrable models \cite{BW,I-PRL}. 
Connection with integrable models was used to derive thermodynamic quantities 
for exclusion statistics in one dimension \cite{FK}.  
Note also that the idea of generalized exclusion principle has been recently 
extended  resulting in statistical weights different from (\ref{weight}) 
\cite{Poly2,CNvD,IIG}.

The original suggestion that
fractional quantum Hall effect quasiparticles obey FES \cite{Hal1}
has since been verified numerically \cite{JC,HXZ} and analytically
\cite{LO}.    
There were also speculations as to the applicability of
FES to other models of condensed matter systems \cite{NW,BBM,BMS}, in which 
context the Thomas-Fermi method was adapted  to study FES \cite{SB}.    

In this paper we discuss the thermodynamics of 
a gas of free (spinless) particles obeying 
FES, deriving
Sommerfeld and virial expansions, extending previous work by
Nayak and Wilczek \cite{NW} on this subject.  In particular, we
derive several new analytic results.  For gases with a constant density
of levels, we obtain an explicit expression for the energy density
as a function of particle density and temperature.  We find in this case that,
at low temperatures, the heat capacity is exactly linear (up to nonanalytic
corrections) and independent of the statistics parameter $g$.  In the
low density expansion, only the second virial coefficient is $g$-dependent.

{\it General Relations} ---  The occupation distribution function derived
from eq. (\ref{weight}) is
\begin{equation}
n(\ve,\mu,T;g)={w\over 1 + (g-1)\, w}\ ,
\label{occ}
\end{equation}
where $w$ satisfies
\begin{equation}
g\ln(1-w) - \ln w = {\ve - \mu\over \kT}\ .
\label{wdef}
\end{equation}
The grand potential is then a sum over the 
single-particle states $i$:
%individual phase space blocks $i$,
%each consisting of $K$ nearby states of energy $\approx \ve_i$:
\begin{equation}
\Omega(T,V,\mu)=\kT\sum_i\ln(1-w_i)\ .
\label{Omega1}
\end{equation}
The  distribution function $n_i$ and 
the single state grand partition function
$\xi_i$, 
related by $n_i=x_i(\partial /\partial x_i)\ln \xi_i$, 
with  $x_i\equiv \exp[(\mu-\varepsilon_i)/\kT]$,
 can be expanded in powers of $x_i$, 
\begin{equation}
\xi_i={1\over 1-w_i}=\sum_{l=0}^\infty P_l\,x_i^l\,,\quad
n_i=\sum_{l=1}^\infty Q_l\,x_i^l \,, 
\label{series}
\end{equation}
where \cite{I2}
\begin{equation}
P_l=\prod_{k=2}^l \left(1-g\,{l\over k}\right) , \quad
Q_l=\prod_{k=1}^{l-1} \left(1-g\,{l\over k}\right) .
\label{PQ}
\end{equation}
These coefficients, which may be introduced as {\it a priori}\ probabilities 
\cite{Poly2},
generally become negative for certain values of $g$.  One can implement a
microscopic formulation of FES which does not rely on negative probabilities
\cite{I3}.

Consider  a gas of free FES particles in $D$ dimensions with dispersion 
$\ve(p)=ap^\sigma$.  The pressure is given by
\begin{equation}
P = -\Omega/V=\Delta\int_0^\infty\!\!\! d\ve\,
\ve^{D/\sigma}\,n(\ve)\ ,
\label{Omega2}
\end{equation}
where $\Delta=a^{-D/\sigma}/(2\sqrt{\pi}\hbar)^D\,\Gamma(1+\half D)$.
Note that $P=(\sigma/D)E/V$, where $E=\sum_i \ve_i\,n_i$ is the energy.
It is convenient to express integrals of the form
\begin{equation}
I[f]=\int_0^\infty\!\!\!d\ve\,f(\ve)\,n(\ve)\ ,
\label{I}
\end{equation}
as ones over the variable $w$,  
\begin{equation}
I[f]=\kT \!\!\! \intl_0^{w(0)}\!\!\!{dw\over 1-w}\,
f \Bigl(\mu+g\kT\ln(1-w) -\kT\ln w\Bigr),   
\label{I-w}\end{equation}
where  $w(0)$ satisfies Eq.~(\ref{wdef}) with $\ve=0$.

{\it Low Temperatures} ---  
At low temperatures, Eq.~(\ref{I-w}) results in the 
Sommerfeld expansion  
%$$
%I[f]=g^{-1}\!\!\!\!\!\!\!\!\intl_{\kT\ln w(0)}^\mu\!\!\!\!\!\!\!\!\!
%d\ve \,f(\ve)
%  +\sum_{l=0}^\infty {(\kT)^{l+1}\over l!}\,C_l(g)\,f^{(l)}(\mu),
%$$
%$$
%C_l(g)=\intl_0^{w(0)}\!\!\!{dw\over 1-w}\Big\{[g\ln(1-w)-\ln w]^l
%-[g\ln(1-w)]^l\Big\}.
%$$
\begin{eqnarray}
I[f]&=&g^{-1}\!\!\!\!\!\!\!\!\intl_{\kT\ln w(0)}^\mu\!\!\!\!\!\!\!\!\!
d\ve \,f(\ve)
+\sum_{l=0}^\infty {(\kT)^{l+1}\over l!}\,C_l(g)\,f^{(l)}(\mu),\nonumber\\
C_l(g)&=&\!\!\intl_0^{w(0)}\!\!\!{dw\over 1-w}\Big\{[g\ln(1-w)-\ln w]^l
-[g\ln(1-w)]^l\Big\} \nonumber .
\end{eqnarray}
Here $w(0)=1-{\cal O}(e^{-\mu/g\kT})$ and can 
be set to unity, up to nonperturbative
corrections
(it is necessary that  $g > 0$).  
Our result for $C_l(g)$ is equivalent to that derived in
Ref.~\cite{NW}, but in its present form we clearly see that $C_l(g)$
is a polynomial of order $(l-1)$ in g, 
\begin{eqnarray}
C_l(g)=&&\sum_{k=0}^{l-1} C_{l,k}\, g^k , \nonumber\\
C_{l,k}=(-1)^{l-k}{l\choose k}\,&&\int_0^1{dw\over w}\,\ln^k(w)
\,\ln^{l-k}(1-w)\ .\label{coefs}
\end{eqnarray}
It is now straightforward to derive the duality relation
\begin{equation}
C_{l,k}=(-1)^{l-1}\,C_{l,l-k-1}
\end{equation}
as well as the results
\begin{eqnarray}
C_0(g)=0,\quad C_1(g)&=&{\pi^2\over 6},\quad C_2(g)=2\,\zeta(3)\,(1-g),\nonumber\\
C_3(g)&=&{\pi^4\over 15}\,(1-\fourth g + g^2) \nonumber\ .
\end{eqnarray}
This proves the conjecture $C_0(g)=0$ (see also Ref.~\cite{Poly2})
and agrees with the numerical estimates of $C_1(\half)$ and $C_2(\half)$
from Ref.~\cite{NW}.

{From} the Sommerfeld expansions for the particle number and the energy, we find 
a chemical potential 
\begin{equation}
\mu=\mu_0 \left\{1-
{\pi^2\over 6}g \left({D\over \sigma}-1\right)\left({\kT\over\mu_0}\right)^2 
+\ldots \right\}  
\end{equation}
and a specific heat 
\begin{equation}
{C_{V,N}\over V}={gDn\over \sigma}\,\left({\kT\over\mu_0}\right)
\left\{{\pi^2\over 3}+
6\,\zeta(3)\,(1-g)\,\left({D\over\sigma}-1\right)
\left({\kT\over\mu_0}\right)+\ldots\right\},
\end{equation}
%\begin{eqnarray}
%{C_{V,N}\over V}&=&{gDn\over \sigma}\,\left({\kT\over\mu_0}\right)
%\left\{{\pi^2\over 3}+\right.\nonumber\\
%&&\left.6\,\zeta(3)\,(1-g)\,\left({D\over\sigma}-1\right)
%\left({\kT\over\mu_0}\right)+\ldots\right\},
%\end{eqnarray}
where $\mu_0=(g\rho/\Delta)^{\sigma/D}$ is the  $T=0$ chemical potential 
and $\rho=N/V$ is the particle density. 

{\it Examples} ---  
When $\ve(p)=p^2/2m$, one obtains a compressibility $\kappa=D/2n\mu_0$
and a sound velocity $v_s=\sqrt{2\mu_0/mD}$.  Restricting further to the
case $D=1$, the sound velocity $v_s=\pi\hbar g n/m$ corresponds to the
pseudo Fermi velocity $v_F= p_F/m$  
($n(p)=1/g$ for $|p|\leq p_F$, otherwise
$n(p)=0$ for $T=0$).  The specific heat is
\begin{equation}
{C_{V,N}\over V}={\pi \kB^2 T\over 3 \hbar v_s}
-{6\zeta(3)\over\pi}(1-g){\kB^3T^2\over m \hbar v_s^3}+\ldots 
\label{C-Suth}\end{equation}
This calculation provides a way to evaluate  thermodynamic quantities
(without restriction only to leading terms)
for the Sutherland model (nonrelativistic particles on a ring of length $L$
interacting via $V(x_i-x_j)=\lambda(\lambda-1)(\pi^2/mL^2)/\sin^2
[\pi(x_i-x_j)/L]$ \cite{Suth}).  Indeed, one may interpret the
thermodynamic Bethe {\it Ansatz\/} equations for the family of solutions
of the Sutherland model with $|\Psi|\propto |x_i-x_j|^\lambda$
as $|x_i-x_j|\to 0$ ($\lambda\geq 0$) as describing an ideal gas
of nonrelativistic pseudoparticles obeying $D=1$ FES with $g=\lambda$,
with corresponding thermodynamic properties \cite{I2,BW,Ha}.  
The leading term of the  low temperature heat capacity 
known for the Sutherland model\cite{Suth} is recovered by the first term in   
(\ref{C-Suth}) (which itself, along with the above sound velocity $v_s$ and 
compressibility $\kappa=1/nmv_s^2$,  
manifests in thermodynamics  a  bosonization of the Sutherland model   
\cite{Hal2,KY}).

{\it Cluster and Virial Expansions} ---  
For the general dispersion $\ve(p)=a\,p^\sigma$,  
expanding Eq.~(\ref{Omega1}) in powers of $x_i$ 
and making summation over $i$, 
we derive a cluster expansion 
\begin{equation}
-{\Omega\over \kT}=
%{P\over \kT}=
\sum_{l=1}^\infty b_l\,z^l\ ,\qquad b_l={Q_l\over l}
Z_1(T/l)\ ,
\end{equation}
where $z=e^{\mu/\kT}$ is the fugacity,  and  
$Z_1(T)=V\,\Delta\,\Gamma(1+D/\sigma)\,
(\kT)^{D/\sigma}$ is the one-particle  partition function
$\sum_i e^{-\varepsilon_i/\kT}$.  One then derives in the usual
way the virial expansion
\begin{eqnarray}
{P\over  \kT}&=&\sum_{l=1}^\infty a_l(T)\,{\rho}^l , \quad 
a_l\equiv \tilde a_l V^{l-1} , \nonumber\\
\tilde a_1&=&1,\quad \tilde a_2=-\bt_2,\quad \tilde a_3=-2\bt_3+4\bt_2^2,\nonumber\\
\tilde a_4&=&-3\bt_4 + 18\bt_3\bt_2-20\bt_2^3,\quad{\rm etc.}\nonumber
\end{eqnarray}
and 
$\bt_l\equiv b_l/b_1^l$.  For example, we find  
\begin{equation}
a_2=2^{-D/\sigma}(g-{\textstyle\half}){V\over Z_1(T)}, \nonumber\\
\end{equation}
with linear dependence on the statistics parameter, and 
\begin{equation}
a_3=\left\{(4^{-D/\sigma}-{\textstyle\frac{2}{3}} 
\cdot 3^{-D/\sigma})
+g(g-1)(4^{1-D/\sigma}-3^{1-D/\sigma})\right\}
\left({V\over Z_1(T)}\right)^2\ .
\end{equation}
%\begin{eqnarray}
%a_3&=&\left\{(4^{-D/\sigma}-{\textstyle\frac{2}{3}} 
%\cdot 3^{-D/\sigma})\right.\nonumber\\
%&&\left.+g(g-1)(4^{1-D/\sigma}-3^{1-D/\sigma})\right\}
%\left({V\over Z_1(T)}\right)^2\ .\nonumber
%\end{eqnarray}
For nonrelativistic particles and $D=1$, these  expressions 
with $g=\lambda$ 
recover 
the virial coefficients for the Sutherland model \cite{dVO2}. 

{\it Gas with Constant Density of States} --- In the cases $D=\sigma$, the
density of states is constant in energy and one  can evaluate the
equation of state exactly.  
The integral (\ref{I-w}) applied for the particle number yields 
$w(0)=1-e^{-\rho/\Delta \kT}$. From Eq.~(\ref{wdef}), we then find 
\begin{equation}
\mu({\rho},T)={g{\rho}\over\Delta}+\kT\ln(1-e^{-{\rho}/\Delta  \kT}). 
\end{equation}
Finally, applying the integral (\ref{I-w}) for the pressure,
%Combining integrals (\ref{I-w}) for the particle number and the energy, 
%Proceeding from Eq.~(\ref{Omega2}),
we obtain
\begin{equation}
%\mu({\rho},T)&=&{g{\rho}\over\Delta}+\kT\ln(1-e^{-{\rho}/\Delta  \kT}),\\
P({\rho},T)={g {\rho}^2\over 2\Delta}+{1\over\Delta}\!\int_0^{\rho}\!\!d\rho'\>
{\rho'\over \exp(\rho'/\Delta \kT)-1}\ .
\label{EOS}
\end{equation}
This result can be expanded to give low temperature and low density series.
Remarkably, for ${\rho}\gg\Delta \kT$, the full perturbative result is
\begin{equation}
P={E\over V}={g{\rho}^2\over 2\Delta}+{\pi^2\over 6}\,\Delta\,(\kT)^2\ ,
\end{equation}
and hence $C_{V,N}/V=\frac13\pi^2\Delta \kB^2 T$ to all orders in $T$.
In the low density limit, we find
\begin{equation}
{P\over \kT}= {\rho} + \frac14(2g-1)\,{{\rho}^2\over\Delta \kT}+\sum_{l=2}^\infty
{\cB_l\over (l+1)!}\,{{\rho}^{l+1}\over(\Delta \kT)^l}\ ,
\label{VirExp}\end{equation}
where $\cB_l$ is the $l^{\rm th}$ Bernoulli number ($\cB_2=\frac{1}{6}$,
$\cB_4=-\frac{1}{30}$, etc.), vanishing for $k$ odd.

As is seen  from Eq.~(\ref{VirExp}), only the second virial coefficient
is $g$-dependent. A similar claim about a virial expansion was made 
in Ref.~\cite{MS}
for  the Calogero model      
(particles with an inverse square interaction in a harmonic potential  
$\frac12 m\omega^2x^2$
on a line) \cite{Cal}, which is a realization of FES and 
has a constant density of states in the quasicontinuous limit
$\hbar\omega\ll \kT$. 
We should note however that the equation of state for the Calogero model, 
because of the presence of a  
harmonic potential,  
cannot be directly related to the 
above calculations for a free gas (see also discussion in 
Refs.~\cite{dVO2,MS2}). We also notice a formal analogy  
between the above thermodynamic properties of free gases with $D=\sigma$ 
and those obtained by Sen and Bhaduri for $D=1$ FES particles in a 
harmonic potential in the Thomas-Fermi approximation \cite{SB}, where the 
density of states is also constant.   

The above results imply that for $D=\sigma $ only $T=0$ quantities  
depend on the statistics parameter.  
The case  $D=\sigma$, in particular, corresponds to a gas of nonrelativistic 
particles in two dimensions, in which case 
$\Delta \kT$ is equal to $\lambda_T^{-2}$,  
where $\lambda_T=\hbar\sqrt{2\pi/m\kT}$ is the thermal de Broglie 
wavelength   
(for the latter case, the virial expansion (\ref{VirExp}) was conjectured 
in Ref.~\cite{I2}). 
It follows that investigations of whether   
FES  occurs in two-dimensional systems (see e.g. Ref.~\cite{NW})  
should refer to $T=0$ properties. 
 
In conclusion, we have derived explicitly a few first terms of both 
low temperature and low density  expansions for 
gases of free spinless particles obeying FES.
For gases with a  constant density of  states in energy, we derived 
the exact equation of state  (\ref{EOS}), which implies that the 
finite-temperature properties are identical to those for a Fermi gas and  
the statistics only appears  in the zero-temperature quantities.

\bigskip

S.B.I. is grateful to J.M. Leinaas, S. Mashkevich, and S. Ouvry for 
stimulating discussions. 
We would like to thank the Centre for Advanced Study (Oslo) for kind 
hospitality and financial support.

\end{document}